\documentclass[a4paper,11pt]{article}
\usepackage{jinstpub} % for details on the use of the package, please see the JINST-author-manual
\usepackage{textgreek}
\usepackage{amsmath}
\usepackage{float}
\usepackage{amsmath}
\usepackage{wrapfig}
\usepackage{subcaption}
\usepackage{graphicx}

\usepackage[backend=biber,
style=numeric,
citestyle=chem-acs]{biblatex}
\title{\boldmath Compositional and morphological study of a Nuragic bronze figurine with neutron diffraction and neutron tomography}

\author[a]{M. Cataldo\note{Corresponding author.}}
\affiliation[a]{INFN Sezione Milano-Bicocca,\\
Piazza della Scienza 3, Milano, Italy}
\author[b]{B. Billeci}
\affiliation[b]{Università degli studi di Sassari,\\
Piazza Università 21, Sassari, Italy}
\author[c]{S. Britto}
\affiliation[c]{ISIS Neutron and Muon Source,\\
Harwell Campus, Didcot, UK}
\author[d]{N. Canu}
\affiliation[d]{Soprintendenza Archeologia, Belle arti e paesaggio per le province di Sassari e Nuoro,\\
Piazza Sant’Agostino 2, Sassari, Italy}
\author[a]{M. Clemenza}
\author[a,c,e,1]{G. Marcucci}
\affiliation[e]{Università degli Studi di Milano-Bicocca,\\
Piazza dell'Ateneo Nuovo 1, Milano, Italy}
\author[c]{A. Scherillo}
\author[b]{V. Sipala}
\author[f]{P. Oliva}
\affiliation[f]{Università di Pisa,\\
Lungarno Pacinotti 43, Pisa, Italy}

\emailAdd{giulia.marcucci@unimib.it}

\abstract{Nuragic figurines are rare and unique examples of the mastery achieved by Sardinian craftsmen in the early Iron Age. These bronze artefacts were most likely cast using the lost wax technique: the shapes were moulded with relative ease, and even complex figures could be represented. However, the manufacturing process was not always a single-step procedure: in some cases, the parts of the model were moulded separately and then assembled together. Therefore, the analytical study of Nuragic bronzes can help to understand the specific casting methods and to evaluate the techniques implemented by Sardinian craftsmen to produce such complex objects. In recent years, Time of Flight Neutron Diffraction (Tof-ND) and Neutron Imaging (NI) have proven to be among the most effective methods for non-invasive studies. Neutron diffraction and neutron imaging provide complementary quantitative and morphological information that can be the key to understanding the casting processes. In this work, the results of the analysis of a bronze figurine will be reported. The statuette represents a Nuragic warrior and was made available by the \textit{Soprintendenza Archeologia Belle Arti e Paesaggio per le province di Sassari e Nuoro} (Sassari, Italy).}


\begin{document}
\maketitle
\flushbottom
\begin{refsection}
    
\section{Introduction}
\label{sec:intro}

\begin{wrapfigure}{r}{0.275\textwidth}
  \begin{center}
    \includegraphics[width=0.2\textwidth]{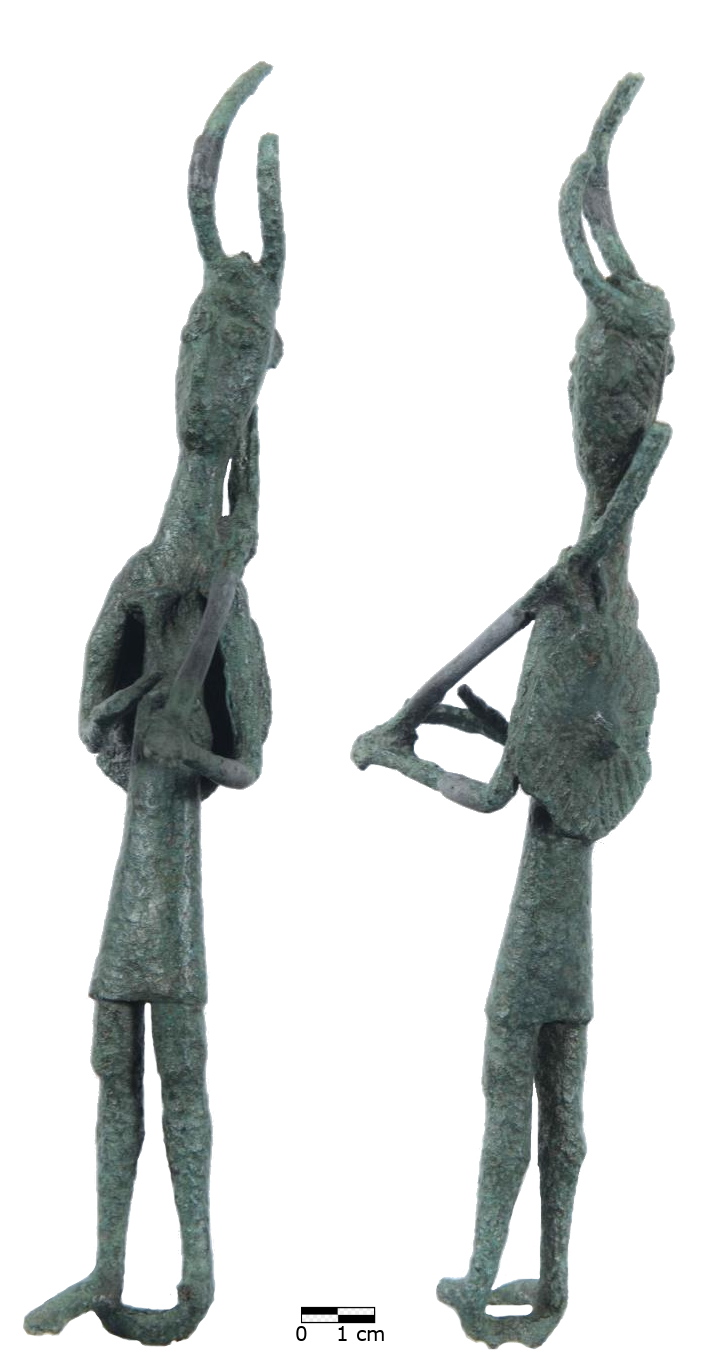}
  \end{center}
  \caption{The warrior statuette. \label{fig:i}}
\end{wrapfigure}

Born in Sardinia in the late Bronze Age, the Nuragic Culture is recognised worldwide for its astonishing artistic and architectural production. From the \textit{Nuraghi} - stone towers that still shape the island landscape to this day - to ceramic, stone and metallic artefacts, Sardinian craftsmen achieved an unmatched level of control of many manufacturing processes. In the late Nuragic period (X-VIII century BCE), the metallic production is characterised by the creation of bronze artefacts of exceptional quality. From objects of everyday use to small statuettes representing complex scenes, the Nuragic culture mastered the art of bronze making. However, a unique trait of the metallic production is that, besides the large number of artefacts, there is very little evidence of the manufacturing methods. Remains of metallic workshops, like furnaces, crucibles or moulds, are very rare in archaeological excavations \cite{uno}. Therefore, the study of these artefacts becomes essential for understanding the production process. In the last few years, neutron-based techniques have emerged as among the most appropriate and effective methods for studying metallic artefacts. Based on the use of a totally non-invasive probe, neutron methods (namely neutron diffraction and imaging) have been applied in many works to obtain qualitative and quantitative bulk information \cite{tre,four}. The results of these experiments have helped scientists and historians to shed light on the manufacturing processes and to understand the history, commercial routes and contaminations of this unique culture. In this work, the results of the study of a small bronze figurine are reported. The statuette represents a lightly armed warrior (Fig. 1), found during the archaeological excavation in \textit{Cuccuru Mudeju}, a small site in the north of Sardinia (\textit{Nughedu San Niccolò}). The sample belongs to the collection of the \textit{Soprintendenza Archeologia Belle Arti e Paesaggio per le province di Sassari e Nuoro} (Sassari, Italy). It is a male figure with a horned helmet and braids, with a sword placed on the left shoulder. A small rope connects the sword with a circular shield with radial motifs, while on the chest is placed a small gamma-shaped hilt dagger, a symbol of adulthood. The right hand of the warrior has four fingers united, a symbol of ritual salutations. Finally, the warrior's open legs end in a small pin, which is believed to be the fusion channel as well as an aid to set the warrior still. The warrior iconography refers to heroic virtues and power, an important aspect of the Nuragic civilisation and most likely served as an offering to Deities. After the excavation, the warrior has been restored in the restoration centre of \textit{Li Piunti} (Sassari, Italy). The right horn, the sword and the left arm were fragmented. The restoration integrated the missing pieces with Araldite, an epoxy resin. In addition, the classic bronze superficial alterations were cleaned, and the warrior is now in a good state of conservation.

\section{Methods}
Neutrons are an invaluable probe for the analysis of archaeological artefacts: they are non-destructive and can deeply penetrate to give access to bulk properties of a material. They can reveal structures at the microscopic scale, provide phase composition, 3D images of the inner parts of the artefacts and elemental and isotopic mapping. For Heritage Science applications, they can be used to answer the most typical archaeometry questions. Neutrons can help researchers to understand the manufacturing methods, the composition, the morphology and the conservation status of a given artefact. For this reason, the warrior was analysed by means of time-of-flight neutron diffraction (Tof-ND) and neutron imaging (NI). Experiments were performed in November 2023 at the ISIS Neutron and Muon Source (UK). Tof-ND was performed at INES Station (Italian Neutron Experimental Station), while NI at the IMAT (Imaging and MATerials science) imaging station. After irradiation, due to the induced activity, the warrior was stored in an active sample cabin for a few days before being cleared for shipment. 

\subsection{Neutron Diffraction}
The INES beam line is a general-purpose time-of-flight neutron diffractometer. The instrument is equipped with \textsuperscript{3}He detectors to detect neutrons, covering a scattering angle range of about 158°. Detectors are placed at a fixed scattering angle: when Bragg’s law is satisfied, the obtained diffraction figures reflect the position of atoms in a solid. With a Tof-ND measurement, it is possible to determine the weight percentages of crystal phases present in the sample, the alloy composition, the texture (i.e, preferred crystallite orientation), the crystallite size and the presence of residual stress (effect of mechanical working). To calculate the phase fractions, Rietveld refinement is used. The warrior was mounted on the INES sample area and measured in eight different areas, with a spot measurement of 25 or 100 mm\textsuperscript{2}. The collected data were normalised with the Mantid software and analysed with the GSAS software through the EXPGUI interface \cite{six}. For a more detailed description of the beamline, see \cite{seven}. 

\subsection{Neutron Imaging}
IMAT is a cold neutron imaging station designed for non-destructive imaging on a broad range of material science areas. Located in ISIS target station 2, the experimental area is equipped with a rotation and translation stage, with a downstream camera for attenuation-based measurements. For a more detailed description of the beam line, see \cite{otto}. The warrior was placed in a custom-made aluminium sample holder and irradiated for a few hours. Tomography measurements were performed using an Andor Zyla sCMOS camera (2048x2048 pixels) coupled to a ZnS/LiF scintillator providing a field of view of $\sim$210x210 mm\textsuperscript{2} and an effective pixel size at the scintillator of $\sim$100$\mu$m. The achievable spatial resolution is influenced by the beam divergence and was optimised by minimising the geometric unsharpness at the scintillator given by: \textit{l\textsubscript{d}/(L/D)}, where \textit{l\textsubscript{d}} is the sample-detector distance, \textit{L} is the distance between the pinhole and the detector, and  \textit{D} is the pinhole aperture diameter. The sample was placed as close to the detector as possible to minimise \textit{l\textsubscript{d}}. With \textit{L} being fixed at $\sim$10 m, a pinhole diameter of 40 mm was selected, corresponding to a collimation ratio (L/D) of 250 and a spatial resolution of $\sim$200$\mu$m. The collected data were processed with the Mantid imaging software and analysed with ImageJ \cite{imageJ}.

%In its current configuration, the instrument supports a field of view of 200x200 mm\textsuperscript{2} for larger samples (with $\sim$200 $\mu$m  spatial resolution), while for smaller samples, a spatial resolution of 50-60$\mu$m can be achieved.

\section{Results and Discussion}
\subsection{Neutron Diffraction}

The phase composition of the eight measured areas is reported in Table 1. The multiphase analysis revealed the presence of a copper-tin alloy characterised by a strong dendritic structure, lead and two secondary phases. To calculate the equivalent tin concentration of the two bronze alpha phases, calibration curves have been used \cite{nine}. As stated by the Vegard law, tin concentration is proportional to the dimension of the lattice cell. The calculated concentration is relatively constant in all the measured areas, averaging to a $\sim$7.5$\pm$0.2 wt\%. Lead, an important element added to facilitate the casting process, has a concentration equal to or lower than 2 wt\%. As the threshold for intentional addition of lead in ancient bronze is around 2 wt\%, in this case, its presence could be attributed both to intentional addition and to the presence of lead in the ores used to refine copper or tin. Where not found, it may not be present or below the detection limit (0.2 wt\%). Finally, secondary phases are cuprite and nantokite, with some traces of chalcocite. Cuprite is generally found in all bronze artefacts, while the presence of nantokite has to be taken into consideration. Nantokite is a highly corrosive phase that can lead to the formation of paratacamite, the hydrate form of chlorides, a very dangerous phase for the metal.

\begin{table}[htbp]
\centering
\caption{Results of the weight concentration phase composition obtained by Rietveld refinement. Five different phases were identified: the two dendritic sub-structures of the cast bronze, lead, cuprite and nantokite (traces of chalcocite were found in the hand and in the left leg). \label{tab:i}}
\smallskip
\begin{tabular}{l|*7c}
\hline
\multicolumn{7}{c}{Phase fractions (wt\%)}\\
\hline
Measured area&Bronze $\alpha$1 &Bronze $\alpha$2&Lead&Cuprite&Nantokite&Av.Tin&J\\
\hline
Hand & 25.8$\pm$1.9 &65.5$\pm$1.9 & - & 5.3$\pm$0.4 & 2.0$\pm$0.2 & 7.9$\pm$0.2 & 1.2\\
Skirt & 37.8$\pm$1.9 &61.0$\pm$1.9 & 0.9$\pm$0.5 & 0.6$\pm$0.1 & 0.5$\pm$0.1 & 7.5$\pm$0.2 & 2.2 \\
Leg & 40.0$\pm$2.4 &55.8$\pm$2.4 & 1.1$\pm$0.7 & 2.1$\pm$0.3 & 0.9$\pm$0.1 & 7.6$\pm$0.2 & 8.3\\
Shield & 35.5$\pm$2.0 &55.2$\pm$2.0 & 1.9$\pm$0.6 & 4.2$\pm$0.3 & 1.2$\pm$0.1 & 7.1$\pm$0.2 & 1.2\\
Neck & 35.4$\pm$0.9 &62.6$\pm$0.9 & 0.9$\pm$0.3 & 0.8$\pm$0.1 & 0.3$\pm$0.1 & 7.7$\pm$0.2 & 3.5\\
Stocco & 36.6$\pm$1.2 &54.1$\pm$1.2 & 1.1$\pm$0.6 & 5.2$\pm$0.3 & 3.0$\pm$0.2 & 7.5$\pm$0.2 & 1.8\\
Horn & 53.3$\pm$1.1 & 32.7$\pm$1.2 & 2.0$\pm$0.8 & 6.7$\pm$0.3 & 5.2$\pm$0.2 & 7.1$\pm$0.2 & 2.0\\
Pin & 28.7$\pm$1.4 & 52.5$\pm$1.4 & - & 8.8$\pm$0.5 & 10.0$\pm$0.4 & 7.7$\pm$0.2 & 1.9\\
\hline
\end{tabular}
\end{table}

What emerges from the analysis of the diffraction pattern is the overall presence of strong texture effects. Texture is the anisotropic distribution of the crystallographic grains, which are oriented at different angles according to the mechanical work and thermal history of the sample. When grains are distributed with anisotropic distribution, or texture, the intensity of the diffraction peak changes from one detector bank to another \cite{ten}. As shown in Figures 2a,b, the copper peaks intensities vary with the scattering angle. From the analysis of the diffraction pattern, it is possible to obtain a parameter that estimates the extent of texture, called texture index (J) and reported in Table 1. This parameter represents the ratio between the volume occupied by all the grains and the volume occupied by isotropic grains. Its value changes from 1 for ideal isotropic grains to infinity for a huge single crystal. 

\begin{figure}[htbp]
\centering
  \begin{subfigure}{0.45\textwidth}
    \includegraphics[width=\linewidth]{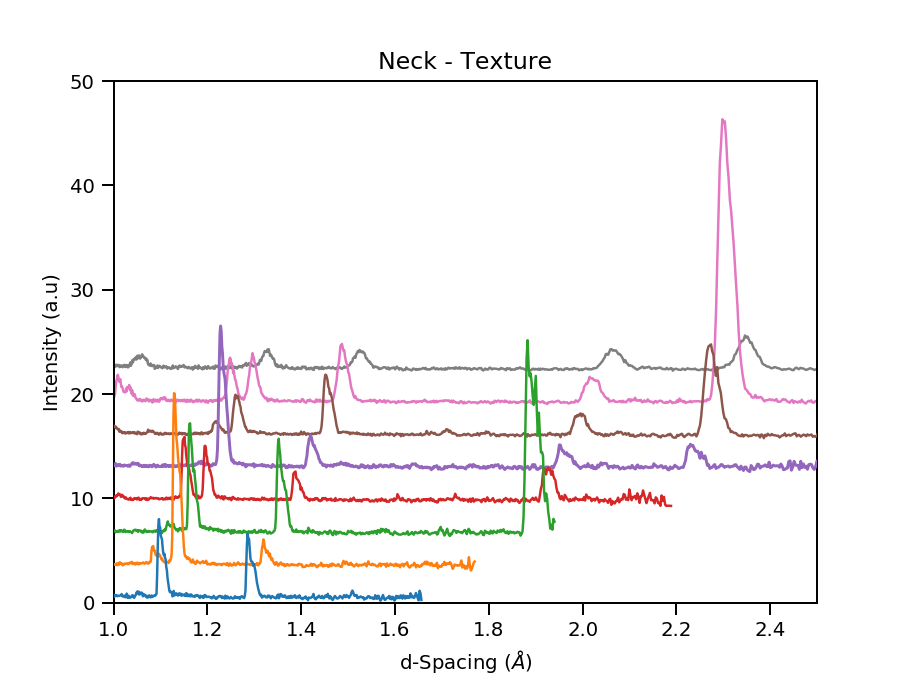}
    \caption{} \label{fig:1a}
  \end{subfigure}%
  \hspace*{\fill} 
  \begin{subfigure}{0.45\textwidth}
    \includegraphics[width=\linewidth]{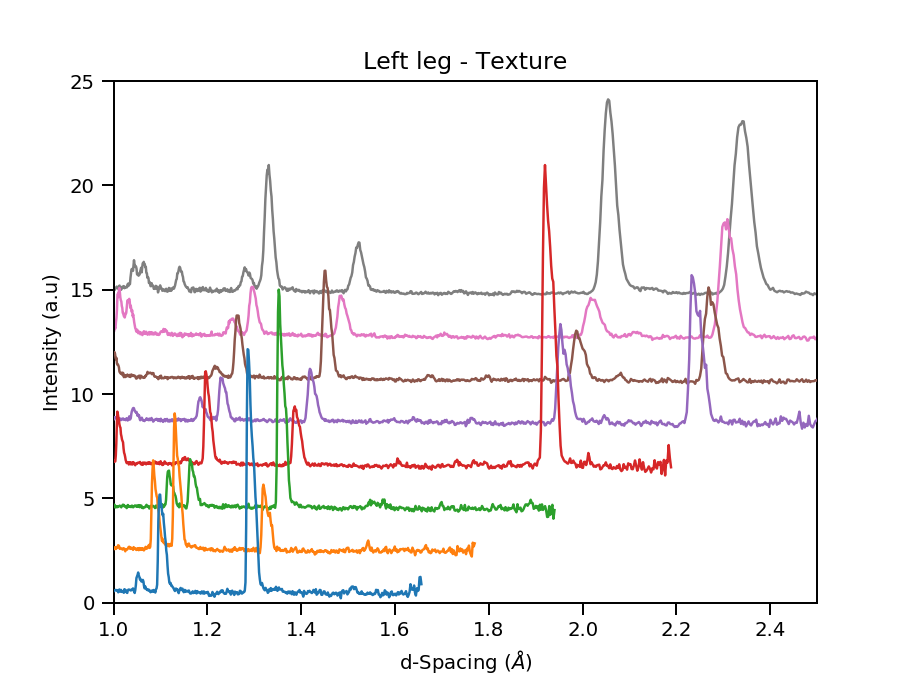}
    \caption{} \label{fig:1B}
  \end{subfigure}%
\caption{Diffraction pattern of some of the areas with the highest texture index. The neck (a) and the left leg (b). Here, it can be seen how the copper peaks at 1.1, 1.27 and 1.89 Å vary their intensity with the variation of scattering angle.}
\end{figure}

\subsection{Neutron Imaging}

\begin{figure}[H]
\centering
\includegraphics[width=0.85\textwidth]{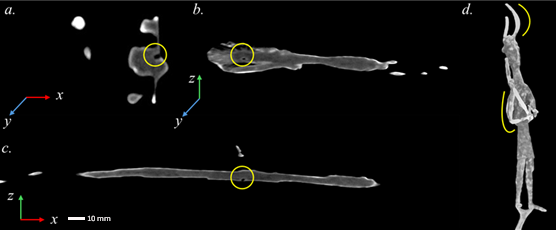}
\caption{Three different slices and a reconstructed 3D projection (d) of the warrior. In the grey scale, the brighter spots correspond to areas rich in elements with a high attenuation coefficient. For the horns and the arms, the responsible is the hydrogen contained in the epoxy resin used for restoration. Circled in yellow, the single pore found in the structure (in the three different projection planes). \label{fig:i}}
\end{figure}

Neutron imaging results are reported in Figure 3. From the analysis of the tomography slices, no discontinuity was observed. Porosity is very limited, except for a small pore in the back of the warrior (circled in yellow). This pore could be proof of the vent-out system. According to archaeological suggestions, the pouring channel could have been the foot pin, and this pore in the back of the warrior could have been used to help ventilation and allow gas to exit the mould. The 3D reconstruction gives a whole vision of the warrior, and helps to identify the details, like the eyes or the dagger placed on the chest. The brighter spots are due to the presence of elements with high attenuation coefficients, likely arising from hydrogen-containing alteration phases. In the case of the horns and the arm of the warrior, this is due to the compound used for the restoration, the epoxy resin “Araldite 2020”. For the thin white layer that surrounds the warrior, instead, this could be either attributed to an alteration layer rich in hydrogen-containing phases or to beam hardening effects \cite{eleven}.

\subsection{Discussion}
The complementary results of the Tof-ND and NI experiments allow to make assumptions about the manufacturing methods of the warrior. It is generally believed that bronze statuettes were made using the lost wax method. The results of the analysis confirm the hypothesis, with no morphological and microstructural characteristics that can suggest a different type of approach. The microstructural analysis, from the characterisation of the alloy peaks, suggests that the warrior underwent a long cooling process – possibly in a preheated mould - as indicated by the presence of the dendritic peaks. The NI results support this hypothesis, with very limited porosity found in the body of the sample. One pore is clearly visible on the back of the warrior. Considering that the foot pin is believed to be the pouring channel, this pore could be evidence of a vent-out system. From the NI results, in addition, no discontinuities were found, suggesting that this sample could be the result of a single cast. The most significant parameter in this analysis is texture, which occurs in all the analysed areas. The presence of texture is evidence of a long cool-down time and cold working, most likely done to adjust the final shape of the statuette. To aid the cold working process, the relatively low tin content (average of 7.5 wt\%) may have been an intentional choice. Finally, as the application of neutron methods to Nuragic bronzes is becoming a standard procedure, it is interesting to compare the results of this work with a recently published paper. In Cantini et al. \cite{twelve}, the results of a neutron imaging study of a bronze statuette are presented. Differently from the findings presented here, the statuette is characterised by the presence of extensive porosities in the body parts and evidence of cast-on of the head, which could have been replaced at a later stage or modified during the casting process. The authors relate the presence of porosity to a lack of control of the manufacturing process. A poor ventilation system prevented the air from flowing out of the mould, especially in the thicker parts of the body, where the porosity is more pronounced. In addition, the tin content of this statuette is higher, with an average value of 10.3 wt\%. The comparison of the results suggests that even if the two artefacts belong to the same category of human-like statuettes, the approach to the manufacturing process has some differences (especially in terms of control of the ventilation system), but the same care was used for the cooling-down process. The subtle contrast, however, can give insight into the different approach of the workshops to the casting of bronze metals, thus helping to characterise this unique production of the island of Sardinia.

\section{Conclusions}
In this work, the use of neutron diffraction and neutron imaging allowed a microstructural and morphological investigation of a Nuragic bronze statuette in a totally non-destructive way. The analysis confirms that the sample is a single piece of cast bronze, with an average tin content of 7.5 wt\%. Though in good conservation state, it has to be considered that some of the measured areas present a high value of nantokite, a copper chloride that could be dangerous for the metal. The results of the analysis, especially if compared to the results of another similar sample, show that the level of control of Sardinian craftsmen was exceptionally high, as they were able to craft very complex figurines with the limited technologies of the time.

\acknowledgments

Experiments at the ISIS Neutron and Muon Source were supported by beamtime allocation RB2220554 and RB2220567 from the Science and Technology Facilities Council \\ 
(data: https://doi.org/10.5286/ISIS.E.RB2220554 \& https://doi.org/10.5286/ISIS.E.RB2220567).

\printbibliography

@book{uno,
    author = "A. Moravetti and E. Alba and L. Foddai",
    title = "La Sardegna Nuragica: storia e materiali" ,
    publisher = "Delfino Editore" ,
    year = "2014"
}

@article{tre,
    author = "A. Fedrigo and F. Grazzi and A. R. Williams and et al" ,
    title = "Extraction of archaeological information from metallic artefacts—A neutron diffraction study on Viking swords",
    journal = "Journal of Archaeological Science: Reports",
    year = "2017",
    doi= "https://doi.org/10.1016/j.jasrep.2017.02.014"
}

@article{four,
    author = "A. Depalmas and M. Cataldo and F. Grazzi and et al." ,
    title = "Neutron-based techniques for archaeometry: characterization of a Sardinian boat model ",
    journal = "Archaeol Anthropol Sci" ,
    year = "2021",
    doi = "https://doi.org/10.1007/s12520-021-01345-w"
}

@article{six,
    author = "Toby B. H",
    title = "EXPGUI, a graphical user interface for GSAS " ,
    journal = "J Appl Crystallography " ,
    year = "2001", 
    doi = "https://doi.org/10.1107/S0021889801002242"
}

@article{seven,
    author = "F. Grazzi and M. Celli and S. Siano and et al" ,
    title = "Preliminary results of the Italian neutron experimental station INES at ISIS: Archaeometric applications" ,
    journal = " Il Nuovo Cimento" ,
    year = "2007", 
    doi = "https://doi.org/10.1393/ncc/i2006-10039-5"
}

@article{otto,
    author = "W. Kockelmann et al. " ,
    title = "Time-of-flight neutron imaging on IMAT@ISIS: a new user facility for materials science ",
    journal = "J Imaging",
    year = "2018",
    doi = "https://doi.org/10.3390/jimaging4030047"
}

@article{nine,
    author = "F. Grazzi and L. Bartoli and S. Siano and et al" ,
    title = "Characterization of copper alloys of archaeometallurgical interest using neutron diffraction: a systematic calibration study",
    journal = "Anal Bioanal Chem",
    year =  "2010",
    doi = "https://doi.org/10.1007/s00216-010-3815-4"
}

@article{ten,
    author = "R. Arletti and L. Cartechini and L.  Rinaldi and Romano et al. ",
    title = "Texture analysis of bronze age axes by neutron diffraction ",
    journal = "Applied Physics A" ,
    year = "2008",
    doi = "https://doi.org/10.1007/s00339-007-4225-0"
}

@article{eleven,
    author = "B. Schillinger and F. Grazzi",
    title = "Artefacts in Neutron CT – Their effects and how to reduce some of them",
    journal = "Physics Procedia ",
    year = "2008",
    doi = "https://doi.org/10.1016/j.phpro.2015.07.034"
}

@article{twelve,
    author = "F. Cantini and O. Sans Planell and A.  Kaestner and et al",
    title = "Non-invasive characterization of the manufacturing process of a Nuragic bronze statuette: a Neutron Imaging study",
    journal = "Journal of Archaeological Science: Reports",
    year = "2024",
    doi = "https://doi.org/10.1016/j.jasrep.2024.104801"
}

@article{imageJ,
    author = "C. Schneider et al",
    title = "NIH Image to ImageJ: 25 years of image analysis",
    journal = "Nat Methods 9",
    year = "2012",
    doi = "https://doi.org/10.1038/nmeth.2089"
}
\end{refsection}
\end{document}